\documentclass{article} 
\usepackage{spconf} 
\title{Hybrid Beamforming: Where Should\\ the Analog Power Amplifiers be Placed?} 
\name{Yasemin Karacora, Ali Kariminezhad and Aydin Sezgin}  
\address{Institute of Digital Communication Systems (DCS), Ruhr University Bochum (RUB)\\ Bochum 44801, Germany \\ email: \{yasemin.karacora, ali.kariminezhad, aydin.sezgin\}@rub.de}

\usepackage[belowskip=-20pt,aboveskip=0pt]{caption}
\usepackage{amsmath}
\usepackage{amssymb}
\usepackage{graphicx}
\usepackage{tikz}
\usepackage{tkz-tab}
\usetikzlibrary{patterns}  
\usepackage{pgfplots}
\usepackage{booktabs}
\usepackage{floatflt}
\usepackage{enumerate}
\usepackage{psfrag}	
\usepackage{array}
\usepackage{multirow,hhline}
\usepackage{exscale}
\usepackage{color}
\usepackage{colortbl}
\usepackage{epsfig}
\usepackage{cite}
\usepackage{amsthm}
\usepackage{array}
\usepackage{subfigure}
\usepackage[font=small]{caption}
\usepackage[latin1]{inputenc} 
\usepackage[T1]{fontenc} 
\usepackage{rotating}
\usepackage{pbox}

\usepackage{algorithm}
\usepackage{algorithmicx}
\usepackage{algpseudocode}
\usepackage{pifont}

\newcommand\norm[1]{\left\lVert#1\right\rVert}

\newcommand{\mat}[1]{\ensuremath{\mathbf{#1}}}
\renewcommand{\vec}[1]{\ensuremath{\mathbf{#1}}}

\newcommand{\VA}{\mat{F}_\mathrm{A}}
\newcommand{\VD}{\mat{F}_\mathrm{D}}
\newcommand{\Vt}{\mat{F}}
\newcommand{\VPS}{\mat{F}_\mathrm{PS}}
\newcommand{\WA}{\mat{W}_\mathrm{A}}
\newcommand{\WD}{\mat{W}_\mathrm{D}}
\newcommand{\Wt}{\mat{W}}
\newcommand{\WPS}{\mat{W}_\mathrm{PS}}

\newtheorem{remark}{Remark}

\usetikzlibrary{shapes,snakes}
\usetikzlibrary{calc}
\usetikzlibrary{patterns}
\usetikzlibrary{decorations.pathmorphing} 
\usetikzlibrary{spy}
\usetikzlibrary{positioning}
\usetikzlibrary{arrows, decorations.markings}

\def\antenna{%
	-- +(0mm,4.0mm) -- +(2.625mm,7.5mm) -- +(-2.625mm,7.5mm) -- +(0mm,4.0mm)
}

\begin{document}


\maketitle
\ninept
\begin{abstract}
In this paper we study the spectral efficiency (SE) of a point-to-point massive multiple-input multiple-output system (P2P-massive MIMO) with limited radio frequency (RF) chains, i.e., analog-to-digital/ digital-to-analog (D2A/A2D) modules, at the transceivers. The resulting architecture is known as hybrid beamforming, where the joint analog and digital beamforming optimization maximizes the SE. We analyze the SE of the system by keeping the number of RF-chains low, but placing analog amplifiers at different paths. Conventional hybrid beamforming architecture uses the amplifiers right after the D2A modules. However, placing them at the phase shifters or at the antennas, can effect the SE of hybrid beamforming. We study the optimal placement of the analog amplifiers and pinpoint the amount of loss in case of misplaced amplifiers.
\end{abstract}


\section{Introduction}
As reported in~\cite{Cisco}, a significant increase in the quality-of-service (QoS) demands of the users is expected in future communication networks. Moreover, due to the appearing technologies such as internet-of-things (IoT), a plethora of users will demand to share the available spectrum. To cope with these phenomena, significant attempts have been made to improve the systems spectral efficiency (SE). For instance, massive MIMO is an outstanding technique to obtain significantly high SE. However, to guarantee the full functionality of massive MIMO, a complicated and costly infrastructure is required. Moreover, the number/size of the antennas can not be made arbitrarily large due to the dimension of the user terminals. Hence, having the massive MIMO architecture within a small implementation area requires operating at significantly high frequencies in order to decrease the antenna array dimensions. This way a large antenna array can provide the functionality of MIMO in small hand-held devices.

Operating at high frequency spectrum, e.g., millimeter-wavelength spectrum (mm-wave) and even in Tera-Hertz regimes provides the feasibility for wideband signal generation, while those spectrum regions are yet idle. Hence, studying the efficiency of massive MIMO in mm-wave spectrum has been the focus of many researchers. For instance, the authors in~\cite{Puglielli2016} study the energy and cost efficiency of mm-wave massive MIMO systems, and the authors in~\cite{Sun2017} study the SE maximization of such systems. 
One of the main bottle-necks of having full-functional massive MIMO in mm-wave spectrum is the requirement to implement a massive number of digital-to-analog/analog-to digital modules (D2A/A2D)-- RF chains-- at the user terminals. With current technologies, these modules are bulky and costly. Hence, despite a large antenna array, a limited number of RF chains are exploited in practical massive MIMO systems~\cite{Puglielli2016}. Using a limited number of RF chains at the user terminals while having a massive antenna array degrades the spectral efficiency of massive MIMO at significantly high signal-to-noise-ratio (SNR). Defining the degrees-of-freedom (DoF) as the approximation to the capacity at high SNR, by limiting the number of RF chains, massive MIMO can only provide a limited DoF independent of the number of antennas. Now, the question is; what is the maximum SE of massive MIMO with an arbitrarily large antenna array and limited number of RF chains? To answer this question, one approach is to separate the digital and analog processing at the transceivers, which results in a hybrid beamforming architecture. Hence, optimal transceiver design is required for this type of beamforming architecture. The authors in \cite{sohrabi2016hybrid} study the optimal analog and digital beamforming design in hybrid P2P-MIMO systems. Moreover, the authors in \cite{utschick2018hybrid} provide an algorithm for obtaining the sub-optimal digital and analog beamforming solutions in cellular downlink. These works consider the analog amplifiers right after the D2A converters.

\textit{Our contribution:} In this paper, we place the analog amplifiers at different paths, namely, at the antennas, and at the phase shifters. This requires a large number of amplifiers due to the large antenna array structure in massive MIMO systems. Explicitly speaking, let $N^{\text {RF}}$ be the number of RF chains and $N$ the number of antennas, we study the spectral efficiency of hybrid beamforming with $N$ and $NN^{\text {RF}}$ number of amplifiers. Exploiting $NN^{\text {RF}}$ amplifiers at the phase shifters, we show the ultimate SE of the hybrid beamforming architecture when the rank of the channel is restricted by $N^{\text {RF}}$.
However, note that when comparing the performance of different hybrid beamforming architectures, we do not take into account any impacts this might have in terms of circuitry.

\textit{Notation:} Throughout the paper, we represent vectors using boldface lower-case letters and matrices using boldface upper-case letters. ${\rm{Tr}}(\bf{A})$, $|{\bf{A}}|$, ${\bf{A}}^{H}$, ${\bf{A}}^{T}$, ${\bf{A}}^{-1}$ represent the trace, determinant, hermitian, transpose and inverse of matrix $\mathbf A$, respectively. ${\bf I}_N$ denotes the identity matrix of size $N$.

\section{System Model}
We consider a point-to-point MIMO system, where the transmitter is equipped with $N$ antennas and $N_t^\mathrm{RF}$ RF chains communicating to a receiver with $M$ antenna elements and $N_r^\mathrm{RF}$ RF chains. The number of data streams to be delivered is denoted by $d$, assuming that $d \leq N_t^\mathrm{RF} \leq N$ and $d \leq N_r^\mathrm{RF} \leq M$.
The hybrid precoding matrix $\Vt$ is hence defined as 
\begin{equation}
    \Vt = \VA \VD,
\end{equation}
in which $\VD \in \mathbb{C}^{N_t^\mathrm{RF} \times d}$ denotes the digital precoder at baseband and $\VA \in \mathbb{C}^{N \times N_t^\mathrm{RF}}$ is the analog precoder realized by phase shifters. Thus, the transmit signal is given by  
\begin{equation}
    \vec{x} = \VA \VD \vec{s},
\end{equation}
in which $\vec{s}\in\mathcal{C}^d$ and $\mathbb{E}\{\|\vec{s}\|^2\} = 1$. Furthermore, the precoder has to fulfill the total transmit power constraint $\mathbb{E} \{ \norm{\vec{x}}^2 \} =  \mathrm{Tr}(\Vt \Vt^H) \leq P_\mathrm{max}$. The received signal is then given by
\begin{equation}
    \vec{y} = \mat{H} \vec{x} + \vec{n},
\end{equation}
where $\mat{H} \in \mathbb{C}^{M \times N}$ denotes the channel matrix and $\vec{n} \sim \mathcal{CN}(\vec{0}, \sigma^2 \mat{I}_M)$ represents the zero-mean additive white Gaussian noise with covariance matrix $\sigma^2 \mat{I}_M$. 

The channel matrix is assumed to follow the Kronecker model given by
\begin{equation}
    \mat{H} = \mat{R}^{1/2} \tilde{\mat{H}}  \mat{S}^{1/2},
    \label{rank_def_channel}
\end{equation}
in which $\mat{R}$ and $\mat{S}$ denote the receive and transmit spatial correlation matrix, respectively~\cite{bolcskei2000performance}. Due to the limited scattering in mm-wave channels we consider a geometric channel model with $L$ paths between transmitter and receiver. Assuming that both the transmitter and the receiver are equipped with uniform linear array (ULA), the matrix $\tilde{\mat{H}}$ is modeled as \cite{sohrabi2016hybrid}
\begin{equation}
\label{geom_channel}
    \tilde{\mat{H}} = \sqrt{\frac{N M}{L}} \sum_{l=1}^{L} \alpha_l \vec{a}_r (\phi_r^l) \vec{a}_t (\phi_t^l)^H.
\end{equation}
Here, $\alpha_l \sim \mathcal{CN} (0,1)$ denotes the complex gain of the $l$-th path, and $\phi_r^l$ and $\phi_t^l$ are the angles of arrival (AoA) and departure (AoD), respectively. Furthermore, $\vec{a}_r (\phi_r^l)$ and $\vec{a}_t (\phi_t^l)$ denote the array response vectors of the receive and transmit antenna array, respectively. For an array with $N_a$ antenna elements and an antenna spacing of half the transmission wavelength, the response vector can be obtained from 
\begin{equation}
    \vec{a} (\phi) = \frac{1}{\sqrt{N_a}} \left[ 1, \dots, e^{j \pi n \sin(\phi)},\dots e^{j \pi (N_a-1) \sin(\phi)} \right]^T
\end{equation}
with $n \in \{0, 1, \dots,N_a-1\}$. 

Defining the receiver analog postcoder $\WA \in \mathbb{C}^{N_r^\mathrm{RF} \times M}$ and the digital postcoder $\WD \in \mathbb{C}^{d \times N_r^\mathrm{RF}}$, the processed signal is given by
\begin{equation}
    \hat{\vec{y}} = \Wt \mat{H} \Vt \vec{s} + \Wt \vec{n},
\end{equation}
in which $\Wt = \WD \WA$.
The achievable rate of the system is then given by \cite{sohrabi2016hybrid}
\begin{equation}
    R = \log_2   \left\lvert \mat{I}_M + \frac{1}{\sigma^2} \Wt^H(\Wt \Wt^H)^{-1}\Wt \mat{H} \Vt \Vt^H \mat{H}^H \right\rvert .
    \label{rate}
\end{equation}

Using a hybrid beamforming structure, we typically have $N_t^\mathrm{RF} \ll N$ and $N_r^\mathrm{RF} \ll M$ due to the high cost of RF chains. Hence, the point-to-point MIMO capacity in \cite{telatar1999capacity} cannot be achieved since the DoF of the system are constrained by the limited number of RF chains. Note that according to \cite{telatar1999capacity}, the optimal fully digital precoder and combiner are derived from the singular value decomposition (SVD) of the channel matrix $\mat{H} = \mat{U}\mat{\Sigma} \mat{V}^H$, in which $\mat{\Sigma}$ is a diagonal matrix containing the singular values of $\mat{H}$ and $\mat{U}$ and $\mat{V}$ are unitary matrices whose columns are the left and right singular vectors of $\mat{H}$, respectively. Using $\mat{V}$ as the precoder and $\mat{U}^H$ as the postcoder and applying water-filling is known to be the optimal scheme that results in $\min (M,N)$ parallel channels.
Similarly, in a hybrid setting with $\min (N_t^\mathrm{RF}, N_r^\mathrm{RF}) < \min (M,N)$, we could decompose the MIMO channel and support up to $\min (N_t^\mathrm{RF}, N_r^\mathrm{RF})$ parallel single-input-single-output (SISO) channels. This is achieved by using the precoder $\mat{V}_c$ as the matrix of $N_t^\mathrm{RF}$ columns of $\mat{V}$ corresponding to the dominant singular values of $\mat{H}$ and the postcoder $\mat{U}_c^H$ being the $N_r^\mathrm{RF}$ dominant left singular vectors of $\mat{H}$. However, in conventional hybrid beamforming architecture, where the analog precoder and postcoder are realized by phase shifters only, it is not possible to realize arbitrary $\mat{V}_c$ and $\mat{U}_c^H$ in general. In this work, we study how the achievable rate in a P2P-MIMO channel with hybrid beamforming can be improved by different placements of analog power amplifiers.

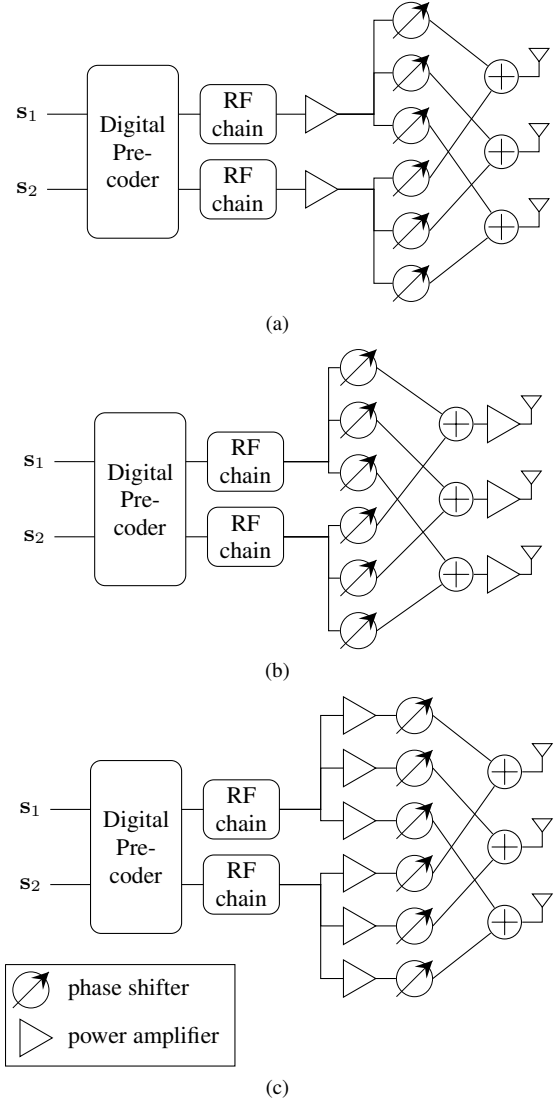
\begin{figure}[tb]
    \centering
    \subfigure[]{\label{fig:standard}\begin{tikzpicture}[scale=0.5,
    box/.style ={draw, rounded corners, align=center, outer sep=0pt}, 
	amp/.style = {regular polygon, regular polygon sides=3,
		draw, fill=white, text width=0.8em,
		inner sep=0mm, outer sep=0mm,
		shape border rotate=-90, scale=0.8},
	ps/.pic={
		\draw[fill=none, inner sep=0,outer sep=0] (0,0) circle (0.3);
		\draw [decoration={markings,mark=at position 1 with
			{\arrow[scale=2.5,>=stealth]{>}}},postaction={decorate}] (-0.3,-0.3) -- (0.3,0.3) ; },
	do path picture/.style={%
		path picture={%
			\pgfpointdiff{\pgfpointanchor{path picture bounding box}{south west}}%
			{\pgfpointanchor{path picture bounding box}{north east}}%
			\pgfgetlastxy\x\y%
			\tikzset{x=\x/2,y=\y/2}%
			#1
		}
	},
	plus/.style={do path picture={    
			\draw [line cap=round] (-3/4,0) -- (3/4,0) (0,-3/4) -- (0,3/4);
		}}
	]

	\node[] (d1) at (-1,1) {$\vec{s}_1$};
	\node[] (d2) at (-1,-1) {$\vec{s}_2$};
	\node[] (dm) at ($(d1)!0.5!(d2)$) {};
	
	\node[box, right of=d1, node distance = 2.8cm, text width=0.8cm](RF1) {RF chain};
	\node[box, right of=d2, node distance = 2.8cm, text width=0.8cm, ](RF2) {RF chain};
	\node[] (RFm) at ($(RF1)!0.5!(RF2)$) {};
	
	\draw[] (d1) -- (RF1);
	\draw[] (d2) -- (RF2);
		
	\node[box, text width=1cm, minimum height=2.3cm, fill = white] (VD) at ($(dm)!0.5!(RFm)$) {Digital Precoder};
	
	\node [amp, right of=RF1, node distance=1.3cm] (amp1) {};
	\node [amp, right of=RF2, node distance=1.3cm] (amp2) {};
	
	\draw[] (RF1) -- (amp1);
	\draw[] (RF2) -- (amp2);
	
	\node[right of=RFm, yshift =0.35cm, node distance=2.3cm, inner sep=0, outer sep=0,scale=0.8] (ps3) {\tikz \draw pic {ps};}; 
	\node[right of=RFm, yshift =-0.35cm, node distance=2.3cm, inner sep=0, outer sep=0, scale=0.8] (ps4) {\tikz \draw pic {ps};}; 
	\node[above of=ps3, node distance=0.7cm, inner sep=0, outer sep=0, scale=0.8] (ps2) {\tikz \draw pic {ps};};
	\node[above of=ps2, node distance=0.7cm, inner sep=0, outer sep=0, scale=0.8] (ps1) {\tikz \draw pic {ps};}; 
	\node[below of=ps4, node distance=0.7cm, inner sep=0, outer sep=0, scale=0.8] (ps5) {\tikz \draw pic {ps};}; 
	\node[below of=ps5, node distance=0.7cm, inner sep=0, outer sep=0, scale=0.8] (ps6) {\tikz \draw pic {ps};}; 
	
	\node[circle, draw, plus, right of=RFm, yshift=1cm, node distance=3.5cm, scale=1.5](plus1) {};
	\node[circle, draw, plus, right of=RFm, node distance=3.5cm, scale=1.5](plus2) {};	
	\node[circle, draw, plus, right of=RFm, yshift=-1cm, node distance=3.5cm, scale=1.5](plus3) {};
	
	\draw (plus1) -- ++(1,0) \antenna;
	\draw (plus2) -- ++(1,0) \antenna;
	\draw (plus3) -- ++(1,0) \antenna;
	
	\draw (ps1.east) -- (plus1);
	\draw (ps2.east) -- (plus2);
	\draw (ps3.east) -- (plus3);
	\draw (ps4.east) -- (plus1);
	\draw (ps5.east) -- (plus2);
	\draw (ps6.east) -- (plus3);
	
	\draw (ps1.west) -- ++(-0.5cm, 0pt) |- (amp1.east);
	\draw (ps2.west) -- ++(-0.5cm, 0pt) |- (amp1.east);
	\draw (ps3.west) -- ++(-0.5cm, 0pt) |- (amp1.east);
	\draw (ps4.west) -- ++(-0.5cm, 0pt) |- (amp2.east);
	\draw (ps5.west) -- ++(-0.5cm, 0pt) |- (amp2.east);
	\draw (ps6.west) -- ++(-0.5cm, 0pt) |- (amp2.east);
	\end{tikzpicture}
	} \subfigure[]{\label{fig:ampAntennas}	\begin{tikzpicture}[scale=0.5,
	box/.style ={draw, rounded corners, align=center, outer sep=0pt}, 
	amp/.style = {regular polygon, regular polygon sides=3,
		draw, fill=white, text width=0.8em,
		inner sep=0mm, outer sep=0mm,
		shape border rotate=-90, scale=0.8},
	ps/.pic={
		\draw[fill=none, inner sep=0,outer sep=0] (0,0) circle (0.3);
		\draw [decoration={markings,mark=at position 1 with
			{\arrow[scale=2.5,>=stealth]{>}}},postaction={decorate}] (-0.3,-0.3) -- (0.3,0.3) ; },
	do path picture/.style={%
		path picture={%
			\pgfpointdiff{\pgfpointanchor{path picture bounding box}{south west}}%
			{\pgfpointanchor{path picture bounding box}{north east}}%
			\pgfgetlastxy\x\y%
			\tikzset{x=\x/2,y=\y/2}%
			#1
		}
	},
	plus/.style={do path picture={    
			\draw [line cap=round] (-3/4,0) -- (3/4,0) (0,-3/4) -- (0,3/4);
		}}
	]

	\node[] (d1) at (-1,1) {$\vec{s}_1$};
	\node[] (d2) at (-1,-1) {$\vec{s}_2$};
	\node[] (dm) at ($(d1)!0.5!(d2)$) {};
	
	\node[box, right of=d1, node distance = 2.8cm, text width=0.8cm](RF1) {RF chain};
	\node[box, right of=d2, node distance = 2.8cm, text width=0.8cm, ](RF2) {RF chain};
	\node[] (RFm) at ($(RF1)!0.5!(RF2)$) {};
	
	\draw[] (d1) -- (RF1);
	\draw[] (d2) -- (RF2);
		
	\node[box, text width=1cm, minimum height=2.3cm, fill = white] (VD) at ($(dm)!0.5!(RFm)$) {Digital Precoder};

	\node[right of=RFm, yshift =0.35cm, node distance=1.5cm, inner sep=0, outer sep=0,scale=0.8] (ps3) {\tikz \draw pic {ps};}; 
	\node[right of=RFm, yshift =-0.35cm, node distance=1.5cm, inner sep=0, outer sep=0, scale=0.8] (ps4) {\tikz \draw pic {ps};}; 
	\node[above of=ps3, node distance=0.7cm, inner sep=0, outer sep=0, scale=0.8] (ps2) {\tikz \draw pic {ps};};
	\node[above of=ps2, node distance=0.7cm, inner sep=0, outer sep=0, scale=0.8] (ps1) {\tikz \draw pic {ps};}; 
	\node[below of=ps4, node distance=0.7cm, inner sep=0, outer sep=0, scale=0.8] (ps5) {\tikz \draw pic {ps};}; 
	\node[below of=ps5, node distance=0.7cm, inner sep=0, outer sep=0, scale=0.8] (ps6) {\tikz \draw pic {ps};}; 
	
	\node[circle, draw, plus, right of=RFm, yshift=1cm, node distance=2.8cm, scale=1.5](plus1) {};
	\node[circle, draw, plus, right of=RFm, node distance=2.8cm, scale=1.5](plus2) {};	
	\node[circle, draw, plus, right of=RFm, yshift=-1cm, node distance=2.8cm, scale=1.5](plus3) {};
	
	\draw (plus1) -- ++(2,0) \antenna;
	\draw (plus2) -- ++(2,0) \antenna;
	\draw (plus3) -- ++(2,0) \antenna;

	\node [amp, right of=plus1, fill=white, node distance=0.7cm] (amp1) {};
	\node [amp, right of=plus2, fill=white, node distance=0.7cm] (amp2) {};
	\node [amp, right of=plus3, fill=white, node distance=0.7cm] (amp3) {};
	
	\draw (ps1.east) -- (plus1);
	\draw (ps2.east) -- (plus2);
	\draw (ps3.east) -- (plus3);
	\draw (ps4.east) -- (plus1);
	\draw (ps5.east) -- (plus2);
	\draw (ps6.east) -- (plus3);
	
	\draw (ps1.west) -- ++(-0.3cm, 0pt) |- (RF1.east);
	\draw (ps2.west) -- ++(-0.3cm, 0pt) |- (RF1.east);
	\draw (ps3.west) -- ++(-0.3cm, 0pt) |- (RF1.east);
	\draw (ps4.west) -- ++(-0.3cm, 0pt) |- (RF2.east);
	\draw (ps5.west) -- ++(-0.3cm, 0pt) |- (RF2.east);
	\draw (ps6.west) -- ++(-0.3cm, 0pt) |- (RF2.east);
	\end{tikzpicture}}
    \subfigure[]{\label{fig:ampPS}\begin{tikzpicture}[scale=0.5,
    box/.style ={draw, rounded corners, align=center, outer sep=0pt}, 
	amp/.style = {regular polygon, regular polygon sides=3,
		draw, fill=white, text width=0.8em,
		inner sep=0mm, outer sep=0mm,
		shape border rotate=-90, scale=0.8},
	ps/.pic={
		\draw[fill=none, inner sep=0,outer sep=0] (0,0) circle (0.3);
		\draw [decoration={markings,mark=at position 1 with
			{\arrow[scale=2.5,>=stealth]{>}}},postaction={decorate}] (-0.3,-0.3) -- (0.3,0.3) ; },
	do path picture/.style={%
		path picture={%
			\pgfpointdiff{\pgfpointanchor{path picture bounding box}{south west}}%
			{\pgfpointanchor{path picture bounding box}{north east}}%
			\pgfgetlastxy\x\y%
			\tikzset{x=\x/2,y=\y/2}%
			#1
		}
	},
	plus/.style={do path picture={    
			\draw [line cap=round] (-3/4,0) -- (3/4,0) (0,-3/4) -- (0,3/4);
		}}
		]

		\node[] (d1) at (-1,1) {$\vec{s}_1$};
		\node[] (d2) at (-1,-1) {$\vec{s}_2$};
		\node[] (dm) at ($(d1)!0.5!(d2)$) {};
		
		\node[box, right of=d1, node distance = 2.8cm, text width=0.8cm](RF1) {RF chain};
		\node[box, right of=d2, node distance = 2.8cm, text width=0.8cm, ](RF2) {RF chain};
		\node[] (RFm) at ($(RF1)!0.5!(RF2)$) {};
		
		\draw[] (d1) -- (RF1);
		\draw[] (d2) -- (RF2);
		
		\node[box, text width=1cm, minimum height=2.3cm, fill = white] (VD) at ($(dm)!0.5!(RFm)$) {Digital Precoder};
		
    	\node[right of=RFm, yshift =0.35cm, node distance=2.3cm, inner sep=0, outer sep=0,scale=0.8] (ps3) {\tikz \draw pic {ps};}; 
    	\node[right of=RFm, yshift =-0.35cm, node distance=2.3cm, inner sep=0, outer sep=0, scale=0.8] (ps4) {\tikz \draw pic {ps};}; 
    	\node[above of=ps3, node distance=0.7cm, inner sep=0, outer sep=0, scale=0.8] (ps2) {\tikz \draw pic {ps};};
    	\node[above of=ps2, node distance=0.7cm, inner sep=0, outer sep=0, scale=0.8] (ps1) {\tikz \draw pic {ps};}; 
    	\node[below of=ps4, node distance=0.7cm, inner sep=0, outer sep=0, scale=0.8] (ps5) {\tikz \draw pic {ps};}; 
    	\node[below of=ps5, node distance=0.7cm, inner sep=0, outer sep=0, scale=0.8] (ps6) {\tikz \draw pic {ps};}; 
    	
		\node[circle, draw, plus, right of=RFm, yshift=1cm, node distance=3.5cm, scale=1.5](plus1) {};
    	\node[circle, draw, plus, right of=RFm, node distance=3.5cm, scale=1.5](plus2) {};	
    	\node[circle, draw, plus, right of=RFm, yshift=-1cm, node distance=3.5cm, scale=1.5](plus3) {};
		
		\draw (plus1) -- ++(1,0) \antenna;
    	\draw (plus2) -- ++(1,0) \antenna;
    	\draw (plus3) -- ++(1,0) \antenna;
		
		\draw (ps1.east) -- (plus1);
		\draw (ps2.east) -- (plus2);
		\draw (ps3.east) -- (plus3);
		\draw (ps4.east) -- (plus1);
		\draw (ps5.east) -- (plus2);
		\draw (ps6.east) -- (plus3);
		
		\draw (ps1.west) -- ++(-2cm, 0pt) |- (RF1.east);
		\draw (ps2.west) -- ++(-2cm, 0pt) |- (RF1.east);
		\draw (ps3.west) -- ++(-2cm, 0pt) |- (RF1.east);
		\draw (ps4.west) -- ++(-2cm, 0pt) |- (RF2.east);
		\draw (ps5.west) -- ++(-2cm, 0pt) |- (RF2.east);
		\draw (ps6.west) -- ++(-2cm, 0pt) |- (RF2.east);
		
		\node [amp, left of=ps1, fill=white] (amp1) {};
		\node [amp, left of=ps2, fill=white] (amp2) {};
		\node [amp, left of=ps3, fill=white] (amp3) {};
		\node [amp, left of=ps4, fill=white] (amp1) {};
		\node [amp, left of=ps5, fill=white] (amp2) {};
		\node [amp, left of=ps6, fill=white] (amp3) {};
		
		\node[below of=d2, scale=0.8, node distance=1.4cm] (ps_legend) {\tikz \draw pic {ps};};
		\node[right of=ps_legend, node distance=1.3cm](ps_text){phase shifter};
		\node[amp, below of=ps_legend ,node distance=0.8cm] (amp_legend) {};
		\node[right of=amp_legend, node distance=1.5cm](amp_text){power amplifier};
		
		\draw [] (ps_legend.north west) -- ++(6.1cm, 0pt) -- ++(0cm, -2.7cm) -- ++(-6.1cm, 0pt) -- ++(0pt, 2.7cm) ;
		\end{tikzpicture}}
    \caption{Three hybrid beamforming transmitter architectures with $N = 3,~d = N_t^\mathrm{RF} = 2$, characterized by the different placement of analog amplifiers. In $(a),~(b)$ and $(c)$, the analog precoder comprises $N_t^\mathrm{RF}$, $N$ and $N N_t^\mathrm{RF}$ amplifiers, respectively.\\}
    \label{fig:HB_architecture}
\end{figure}

As depicted in Figure \ref{fig:HB_architecture}, three hybrid beamforming precoder and combiner structures are studied in this work. In the first structure, $N_t^\mathrm{RF}$ power amplifiers are placed at the RF chains. The second architecture contains $N$ amplifiers, one placed at each transmit antenna. Finally, in the third scheme, the amplification takes place at the analog phase shifters, requiring $N N_t^\mathrm{RF}$ amplifiers. In all three cases, the corresponding hybrid combiner at the receiver is structured in the same way as the precoder. The second and third architecture are considered in order to realize a more accurate approximation of the optimal beamformers $\mat{V}_c$ and $\mat{U}_c^H$.
In the following, the precoding and postcoding schemes are presented for these three hybrid architectures. For the rest of the paper, assume that $N_t^\mathrm{RF} = N_r^\mathrm{RF} = N^\mathrm{RF}$.

\subsection{Amplifiers at the RF chains}
\label{sec_A}
First, consider the architecture shown in Figure \ref{fig:standard}, in which analog power amplifiers are placed at the RF chains. 
We use the scheme presented in \cite[Algorithms~1,2]{sohrabi2016hybrid} to determine the hybrid precoder and combiner. Let $\VPS$ and $\WPS$ be phase shifter matrices for the precoder and postcoder, respectively, with the constraint on the matrix elements $\lvert\VPS(i,j)\rvert = \lvert\WPS(i,j)\rvert = 1 \quad \forall i,j$. Note that $\VA$ and $\WA$ are realized by phase shifters only. Thus, we can write $\VA = \VPS$ and $\WA = \WPS$.
First, $\VPS$ is determined by applying \cite[Algorithm~1]{sohrabi2016hybrid}. Then, the digital precoder is obtained as \cite{sohrabi2016hybrid}
\begin{equation}
    \VD = (\VPS^H \VPS)^{-1/2} \mat{U}_e \mat{\Gamma}_e,
\end{equation}
in which $\mat{U}_e$ is the set of right singular vectors of $\mat{H} \VPS (\VPS^H \VPS)^{-1/2}$ and $\mat{\Gamma}_e = \mathrm{diag}(\sqrt{p_1},\dots,\sqrt{p_d})$ with $p_i$ denoting the power allocated to the $i$-th data stream by water-filling. Similarly, $\WD$ is given by\cite{sohrabi2016hybrid}
\begin{align}
    \WD &= \mat{J}^{-1} \WPS^H \mat{H} \VPS \VD,\\
    \mat{J} &= \WPS^H \mat{H} \VPS \VD \VD^H \VPS^H \mat{H}^H \WPS + \sigma^2 \WPS^H \WPS.
\end{align}
where $\WPS$ follows~\cite[Algorithm~1]{sohrabi2016hybrid}. In what follows, we propose a scheme where the amplifiers are placed at the antennas.

\subsection{Amplifiers at the antennas}
\label{sec_B}
In this scheme, a variable gain amplifier (VGA) is placed at each antenna. The precoding matrix $\Vt = \VA \VD$ is then modeled with the analog precoder
\begin{equation}
    \VA = \mat{B}_\mathrm{t} \VPS 
\end{equation}
in which $\mat{B}_\mathrm{t} = \mathrm{diag} (\beta^\mathrm{t}_{1}, \dots, \beta^\mathrm{t}_{N})$ and $\beta^\mathrm{t}_{k}$ denotes the amplifier gain at the $k$-th transmit antenna. To determine $\VPS$ and $\VD$ the previously described algorithm from \cite{sohrabi2016hybrid} is applied. However, as the power amplification now takes place at the antennas, we define the digital precoder as $\VD = \frac{1}{\sqrt{P_\mathrm{max}}}(\VA^H \VA)^{-1/2} \mat{U}_e \mat{\Gamma}_e$. 

Note that the water-filling matrix $\mat{\Gamma}_e$ is scaling the columns of the precoding matrix $\Vt$. This is because when postmultiplying a matrix by a diagonal matrix, each column of the original matrix is multiplied by the respective diagonal element of the diagonal matrix. In contrast, when premultiplying the precoding matrix by the diagonal matrix $\mat{B}_\mathrm{t}$, the $i$-th row of the precoding matrix is multiplied by $\beta^\mathrm{t}_i,~i = 1,\dots N$.
Recall that the optimal $\Vt$ should be equal to $\mat{V}_c\boldsymbol{\Gamma}_e$ which is not feasible by conventional hybrid beamforming. Further note that while the columns of $\mat{V}_c$ are right singular vectors of $\mat{H}$ and thus have unit norm, $\mat{V}_c$ does not have unit row norm for $N^{RF} < N$. We propose a scheme (see Algorithm \ref{alg:the_alg}) that besides scaling the columns of $\Vt$ through water-filling, also takes advantage of the power amplifiers at the antennas that enable us to independently scale the rows of $\Vt$ in order to make it closer to $\mat{V}_c$. In particular, $\mat{B}_\mathrm{t}$ is determined such that the norm of each row of $\Vt$ becomes approximately equal to the corresponding row norm of $\mat{V}_c$, scaled by $\sqrt{P_\mathrm{max}}$. This is implemented in Algorithm \ref{alg:the_alg}, line 4, where $\vec{v}_{\mathrm{c},k}$ denotes the $k$-th row of $\mat{V}_c$. The amplifier gains $\beta^\mathrm{t}_k$ are successively determined in that manner as long as the total transmit power of the first $k$ antennas $P_k$ does not exceed $P_\mathrm{max}$.

\begin{algorithm}[htb]
\caption{Design of $\mat{B}_\mathrm{t}$}
  \begin{algorithmic}[1]
    \Require $\VD,~\VPS,~P_\mathrm{max},~\mat{V}_c$
    \State Initialize $P_k = 0,~k=0$
    \State $(a_1,\dots,a_N) = \mathrm{diag}(\VPS \VD \VD^H \VPS^H)$
    \While {$P_k < P_\mathrm{max}$}
    \State $\beta^\mathrm{t}_{k} = \sqrt{\frac{N}{N^\mathrm{RF}} P_\mathrm{max}} \norm{\vec{v}_{\mathrm{c},k}}$
    \State $P_k = \sum_{i=1}^{k} {\beta^\mathrm{t}_{i}}^2 a_i$
    \State $k \gets k+1$
    \EndWhile
    \State $\beta^\mathrm{t}_{k-1} = \sqrt{\frac{P_\mathrm{max} - \sum_{j=2}^{k-2}{\beta^\mathrm{t}_{j}}^2 a_j}{a_{k-1}}}$
    \State $\beta^\mathrm{t}_i = 0, \quad i=k,\dots N$
    \State $\mat{B}_\mathrm{t} = \mathrm{diag}(\beta^\mathrm{t}_1,\dots,\beta^\mathrm{t}_N)$
    \end{algorithmic}
    \label{alg:the_alg}
\end{algorithm}

At the receiver, each antenna is equipped with a VGA as well. The amplifier gain is determined in a similar fashion:
\begin{align}
    \mat{B}_\mathrm{r} &= \mathrm{diag}(\beta^\mathrm{r}_1,\dots,\beta^\mathrm{r}_M)\\
    \beta^\mathrm{r}_k &= \frac{1}{\sqrt{N^\mathrm{RF}}} \norm{\vec{u}_{\mathrm{c},k}},
\end{align}
in which $\vec{u}_{\mathrm{c},k}$ is the $k$-th row of $\mat{U}_\mathrm{c}$. After determining $\WD$ and $\WPS$ as described in section \ref{sec_A} using the algorithm from \cite{sohrabi2016hybrid}, the overall postcoding matrix is thus given by
\begin{equation}
    \Wt = \WD \WPS \mat{B}_\mathrm{r}.
\end{equation}

\begin{remark}
As previously mentioned, $\Vt$ is designed in a way such that the norm of its $k$-th row is approximately equal to the corresponding row norm of $\mat{V}_c$ scaled by $\sqrt{P_\mathrm{max}}$. Defining $\vec{f}_k$ and $\vec{f}_{\mathrm{PS},k}$ as the $k$-th row of $\Vt$ and $\VPS$, respectively, we have
\begin{equation}
    \begin{split}
    \norm{\vec{f}_k} &= \beta^\mathrm{t}_k \norm{\vec{f}_{\mathrm{PS},k} \VD}\\
    &=\beta^\mathrm{t}_k \frac{1}{\sqrt{P_\mathrm{max}}} \norm{\vec{f}_{\mathrm{PS},k} (\VPS^H \VPS)^{-1/2} \mat{U}_e \mat{\Gamma}_e}\\
    &\underset{(a)}{\approx} \beta^\mathrm{t}_k \frac{1}{\sqrt{N P_\mathrm{max}}} \norm{\vec{f}_{\mathrm{PS},k} \mat{U}_e \mat{\Gamma}_e}\\
    &\underset{(b)}{\approx} \beta^\mathrm{t}_k \sqrt{\frac{N^\mathrm{RF}}{N}} = \sqrt{P_\mathrm{max}} \norm{\vec{v}_{\mathrm{c},k}}.
    \end{split}
\end{equation}
Here, step (a) holds because $\VPS^H \VPS \approx N \mat{I}_{N^\mathrm{RF}}$ with high probability assuming that $N \gg N^\mathrm{RF}$ \cite{sohrabi2016hybrid}. Step (b) is valid since all elements of $\VPS$ have unit modulus and $\mat{U}_e$ is a unitary matrix. Note that the same holds for the postcoding matrix, except that we do not have a power constraint at the receiver. 
\end{remark}

\subsection{Amplifiers at the phase shifters}
\label{sec_C}
By placing VGAs directly in the phase shifter branches the analog precoder matrix $\VA$ is not constrained to unit modulus elements anymore. This enables an implementation of $\mat{V}_c$ that is only restricted by the precision of the VGAs. Hence, the precoders can be designed with $\VA = \mat{V}_c \mat{\Gamma}_e$ and $\VD = \mat{I}_\mathrm{N^\mathrm{RF}}$. Similarly, the postcoding matrices at the receiver are $\WA = \mat{U}_c^H $ and $\WD = \mat{I}_\mathrm{N^\mathrm{RF}}$.

\begin{remark}
In order to have a fair comparison of the three proposed schemes, we have to ensure that the actual transmit power is equal in all three cases. It can be easily verified that the actual total transmit power in both schemes presented in \ref{sec_A} and \ref{sec_C} is $\mathrm{Tr}(\mat{\Gamma}_e \mat{\Gamma}_e^H)$, which is nearly equal to $P_\mathrm{max}$ to within some small tolerance due to water-filling algorithm. In the scheme described in \ref{sec_B}, the transmit power can be written as
\begin{equation}
\begin{split}
    \mathbb{E} \{ \norm{\vec{x}}^2 \} &=  \mathrm{Tr}(\Vt \Vt^H) = \mathrm{Tr}(\mat{B}_\mathrm{t} \VPS \VD \VD^H \VPS^H \mat{B}_\mathrm{t}^H)\\
    &= \mathrm{Tr} \left( \mat{B}_\mathrm{t}^H \mat{B}_\mathrm{t} \VPS \VD \VD^H \VPS^H \right)
    = \sum_{i=1}^{N} {\beta^\mathrm{t}_i}^2 a_i,
\end{split}
\end{equation}
while $a_i$ denotes the $i$-th diagonal element of $\VPS \VD \VD^H \VPS^H$. 
The statement in Algorithm \ref{alg:the_alg}, line 8 guarantees that the total transmit power is always equal to $P_\mathrm{max}$.
\end{remark}

\section{Numerical Results}
The performances of the three hybrid beamforming schemes presented in the previous section are simulated to compare the average spectral efficiency to the one achieved by fully digital beamforming. For the simulation, we consider a P2P-MIMO system with $M = 8$ receive antennas and a variable number of transmit antennas. For the hybrid beamforming schemes we assume that $N_t^\mathrm{RF} = N_r^\mathrm{RF} = 2$. In our simulations, we consider the two following channel structures
\begin{enumerate}
    \item full-rank channel with $\mathbf{S}^{1/2}=\mathbf{I}_N$ and $\mathbf{R}^{1/2}=\mathbf{I}_M$ 
    \item rank-deficient channel, with  $\mathbf{S}^{1/2}=\mathbf{I}_N$ and $\mat{R}^{1/2} = [\vec{r}_1 \dots \vec{r}_M]^T$
\end{enumerate}
where $(\vec{r}_1, \dots, \vec{r}_{N^\mathrm{RF}})$ are independent unit norm random vectors and the remaining $(\vec{r}_{N^\mathrm{RF}+1}, \dots, \vec{r}_M)$ are linear combinations of the first $N^\mathrm{RF}$ vectors, i.e. $\vec{r}_i = \frac{\sum_{j=1}^{N^\mathrm{RF}} \alpha_j \vec{r}_j} { \norm{\sum_{j=1}^{N^\mathrm{RF}} \alpha_j \vec{r}_j}}, \quad i = N^\mathrm{RF}+1,\dots,M$. This leads to a channel matrix $\mat{{H}}$ of rank $N^\mathrm{RF}$. Note that rank deficiency occurs naturally in mm-wave channels due to the low scattering and high path loss \cite{xie2016overview}. The insight we get from considering the rank-deficient channel is based on the fact that the SE of fully digital structure can be achieved by hybrid beamforming only if $\textrm{rank}(\mathbf{H})=N^{\textrm{RF}}$ and the amplifiers are optimally placed. Otherwise, fully digital beamforming has both SNR and DoF gains, which can not be offered by the hybrid beamforming structure. Hence, the ultimate performance of hybrid beamforming can be guaranteed by the optimal placement of the amplifiers.

In the geometric channel model in~\eqref{geom_channel}, we consider the number of paths to be $L=15$, and the AoA and AoD to be uniformly distributed on the interval $[0,2\pi]$. 
The average spectral efficiency is obtained from \eqref{rate} for a SNR of -5dB averaging over 100 channel realizations.
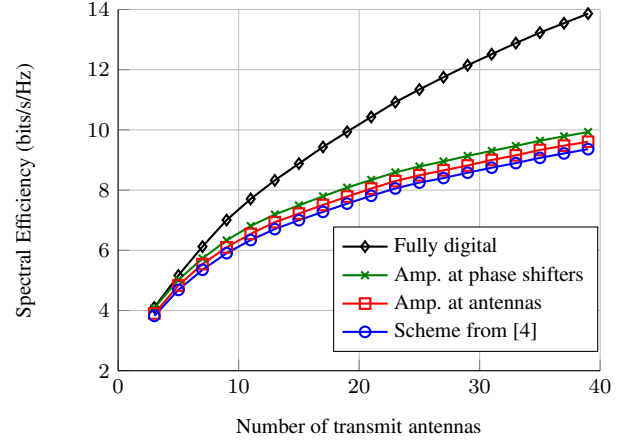
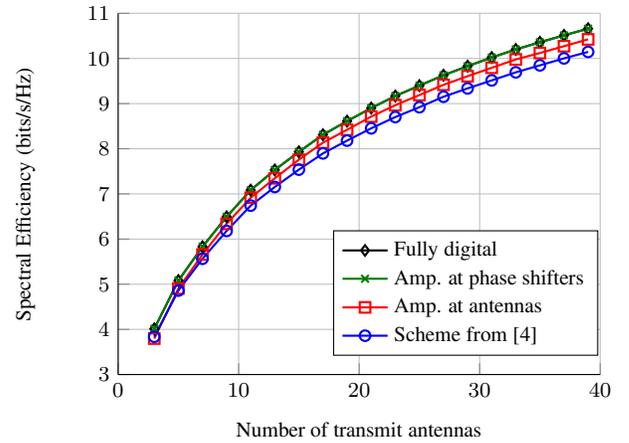
\begin{figure}[htb]
    \centering
    \subfigure[Full rank channel]{\label{fig:full_rank}
%
%
\definecolor{mycolor1}{rgb}{0.00000,0.49804,0.00000}%
\pgfplotsset{
    tick label style = {font = {\fontsize{8 pt}{8 pt}\selectfont}},
  }

\begin{tikzpicture}

\begin{axis}[%
width=0.36\textwidth,
height=0.27\textwidth,
at={(0.853in,0.461in)},
scale only axis,
xmin=0,
xmax=40,
xtick={ 0, 10, 20, 30, 40},
ytick={2,4,6,8,10,12,14},
xlabel style={font={\fontsize{8 pt}{9 pt}\selectfont}},
xlabel={Number of transmit antennas},
ymin=2,
ymax=14,
ylabel style={font={\fontsize{8 pt}{9 pt}\selectfont}},
ylabel={Spectral Efficiency (bits/s/Hz)},
axis background/.style={fill=white},
axis x line*=bottom,
axis y line*=left,
xmajorgrids,
ymajorgrids,
legend style={at={(0.996,0.052)}, anchor=south east, legend cell align=left, align=left, draw=black, font={\fontsize{8 pt}{9 pt}\selectfont}}
]
\addplot [color=black, line width=0.8pt, mark=diamond, mark options={solid, black}]
  table[row sep=crcr]{%
3	4.10063430425771\\
5	5.16427076989133\\
7	6.12209542522659\\
9	7.00498906539452\\
11	7.70496341650452\\
13	8.32294983338852\\
15	8.87930825762192\\
17	9.43255853835234\\
19	9.93577011821186\\
21	10.4300474115958\\
23	10.919294789598\\
25	11.3426458140941\\
27	11.7486025377253\\
29	12.1462316803326\\
31	12.5137717399123\\
33	12.8819381629852\\
35	13.2299647910957\\
37	13.5447054022175\\
39	13.8617987362309\\
};
\addlegendentry{Fully digital}

\addplot [color=mycolor1, line width=0.8pt, mark=x, mark options={solid, mycolor1}]
  table[row sep=crcr]{%
3	4.084233167514\\
5	5.02050811168987\\
7	5.73106922977631\\
9	6.32911997966534\\
11	6.80793546150171\\
13	7.18852155365936\\
15	7.49403657568163\\
17	7.79221475892024\\
19	8.08193295225606\\
21	8.33870123088616\\
23	8.58806067998462\\
25	8.78380344286652\\
27	8.95242507465197\\
29	9.13481919834501\\
31	9.30320894269607\\
33	9.46640045669676\\
35	9.6411739490921\\
37	9.78683478337563\\
39	9.92869922892671\\
};
\addlegendentry{Amp. at phase shifters}

\addplot [color=red, line width=0.8pt, mark=square, mark options={solid, red}]
  table[row sep=crcr]{%
3	3.90151366878609\\
5	4.83056097908454\\
7	5.54316829108463\\
9	6.09843592889948\\
11	6.54644918672973\\
13	6.93428230890232\\
15	7.22057565755117\\
17	7.50630121173416\\
19	7.78436838583785\\
21	8.04970130690555\\
23	8.29763588692461\\
25	8.4909931397778\\
27	8.65675360110429\\
29	8.81704781986711\\
31	9.00067951156034\\
33	9.15369540176815\\
35	9.33329134151565\\
37	9.47596507117871\\
39	9.6048709422171\\
};
\addlegendentry{Amp. at antennas}

\addplot [color=blue, line width=0.8pt, mark=o, mark options={solid, blue}]
  table[row sep=crcr]{%
3	3.82104359084536\\
5	4.68448852947306\\
7	5.35216854172239\\
9	5.90462144093022\\
11	6.33977688408094\\
13	6.70424535948987\\
15	7.00022915855255\\
17	7.27613568553881\\
19	7.55077197762538\\
21	7.81023897277026\\
23	8.05426559599128\\
25	8.24835668881608\\
27	8.40065277984494\\
29	8.57659835114842\\
31	8.74448671027728\\
33	8.89618163404853\\
35	9.0700259414935\\
37	9.218030156391\\
39	9.35954945560984\\
};
\addlegendentry{Scheme from \cite{sohrabi2016hybrid}}

\end{axis}
\end{tikzpicture}%
}
    \subfigure[Rank-deficient channel ($\textrm{rank}(\mathbf{{H}})= N^\mathrm{RF}$)]{\label{fig:rank_def}
%
%
\definecolor{mycolor1}{rgb}{0.00000,0.49804,0.00000}%

\pgfplotsset{
    tick label style = {font = {\fontsize{8 pt}{8 pt}\selectfont}},
  }
\begin{tikzpicture}

\begin{axis}[%
width=0.36\textwidth,
height=0.27\textwidth,
at={(0.853in,0.461in)},
scale only axis,
xmin=0,
xmax=40,
xtick={ 0, 10, 20, 30, 40},
ytick={3,4,5,6,7,8,9,10,11},
xlabel style={font={\fontsize{8 pt}{9 pt}\selectfont}},
xlabel={Number of transmit antennas},
ymin=3,
ymax=11,
ylabel style={font={\fontsize{8 pt}{9 pt}\selectfont}},
ylabel={Spectral Efficiency (bits/s/Hz)},
axis background/.style={fill=white},
axis x line*=bottom,
axis y line*=left,
xmajorgrids,
ymajorgrids,
legend style={at={(0.996,0.052)}, anchor=south east, legend cell align=left, align=left, font={\fontsize{8 pt}{9 pt}\selectfont}}
]
\addplot [color=black, line width=0.8pt, mark=diamond, mark options={solid, black}]
  table[row sep=crcr]{%
3	4.01194956195382\\
5	5.08132599689854\\
7	5.83081210959684\\
9	6.493025157568\\
11	7.08312606555685\\
13	7.53070799624607\\
15	7.93223179751669\\
17	8.3104120568999\\
19	8.61183323622644\\
21	8.90126844803605\\
23	9.16816540416962\\
25	9.39788256679069\\
27	9.62792448252833\\
29	9.82863568231688\\
31	10.0207255452488\\
33	10.1994295828803\\
35	10.358767286113\\
37	10.5120862612823\\
39	10.6593100543833\\
};
\addlegendentry{Fully digital}

\addplot [color=mycolor1, line width=0.8pt, mark=x, mark options={solid, mycolor1}]
  table[row sep=crcr]{%
3	4.01194956187587\\
5	5.08132599673831\\
7	5.83081210940268\\
9	6.49302515735481\\
11	7.08312606533992\\
13	7.53070799602261\\
15	7.93223179729388\\
17	8.31041205667823\\
19	8.61183323600432\\
21	8.90126844781225\\
23	9.16816540394629\\
25	9.39788256656781\\
27	9.62792448230583\\
29	9.82863568209465\\
31	10.0207255450268\\
33	10.1994295826585\\
35	10.3587672858913\\
37	10.5120862610608\\
39	10.6593100541619\\
};
\addlegendentry{Amp. at phase shifters}

\addplot [color=red, line width=0.8pt, mark=square, mark options={solid, red}]
  table[row sep=crcr]{%
3	3.79599903693684\\
5	4.904450703365\\
7	5.65863346996296\\
9	6.33705406456251\\
11	6.91588241152235\\
13	7.34559803345259\\
15	7.75941522310589\\
17	8.13495640865515\\
19	8.42042747822793\\
21	8.71217242074243\\
23	8.96444838444496\\
25	9.18958686481719\\
27	9.41603721684337\\
29	9.60654886278142\\
31	9.79307911112405\\
33	9.97815389415932\\
35	10.1201070236716\\
37	10.2714015404438\\
39	10.4224679570509\\
};
\addlegendentry{Amp. at antennas}

\addplot [color=blue, line width=0.8pt, mark=o, mark options={solid, blue}]
  table[row sep=crcr]{%
3	3.83631053011543\\
5	4.85962029024538\\
7	5.56368407068796\\
9	6.17958715075155\\
11	6.74064879754421\\
13	7.15376650392427\\
15	7.53803572903539\\
17	7.89750800104744\\
19	8.18060686976023\\
21	8.45255817248872\\
23	8.70238963214386\\
25	8.92033979452793\\
27	9.15214585182692\\
29	9.33669873072762\\
31	9.51384387331584\\
33	9.69264740834551\\
35	9.84711250839091\\
37	10.0015039166399\\
39	10.1452767113036\\
};
\addlegendentry{Scheme from \cite{sohrabi2016hybrid}}

\end{axis}
\end{tikzpicture}%
}
    \caption{Average SE as a function of the number of transmit antennas in a P2P-MIMO system with three different hybrid beamforming architectures compared to fully digital beamforming.\\}
    \label{fig:sim_results1}
\end{figure}
Fig.~\ref{fig:sim_results1} shows the average SE achieved by each of the three hybrid beamforming schemes as well as by fully digital beamforming for both the full rank channel and the rank-deficient channel. The scheme from \cite{sohrabi2016hybrid} is outperformed by the two other hybrid beamforming architectures from Fig.~\ref{fig:ampAntennas} and \ref{fig:ampPS}. 
Note that the increase in SE that comes from different amplifier placement is only due to SNR gains, since the DoF is restricted to the number of RF chains in all three hybrid setups.
Furthermore, the performance increases when more amplifiers are deployed, i.e., the scheme using $N N^\mathrm{RF}$ amplifiers at the phase shifters achieves a better SE than the one with $N$ amplifiers at the antennas. 
With fully digital beamforming, however, having a full rank channel the DoF is given by $\min \{N,M\}$, as opposed to $N^\mathrm{RF}$ in the hybrid setting. This is why in Fig.~\ref{fig:full_rank} the gap between the SE achieved by fully digital and hybrid beamforming schemes increases for a large number of transmit antennas $N$.
This is not the case when using the rank-deficient channel model (see Fig.~\ref{fig:rank_def}).
 Since the rank of the channel $\mat{\tilde{H}}$ is equal to the number of RF chains, the DoF is the same in all four setups. Furthermore, the SE achieved when using the scheme from \ref{sec_C} coincides with the SE obtained from fully digital beamforming.

\section{Conclusion}
We studied a P2P-MIMO system with a hybrid beamforming architecture at the transmitter and the receiver. We compared the performance of three different schemes that are characterized by the number and placement of analog amplifiers. In particular, starting from the conventional hybrid architecture and the scheme from \cite{sohrabi2016hybrid} requiring $N^\mathrm{RF}$ amplifiers, we increased the number of amplifiers to $N$ and $N N^\mathrm{RF}$ placing them at the antennas and at the phase shifters, respectively. Proposing an algorithm to determine the amplifier gains, we showed that the spectral efficiency could be improved due to the additional amplifiers. Furthermore, in a rank-deficient channel with rank $N^\mathrm{RF}$ the architecture with amplifiers placed in each phase shifter branch achieves the optimal SE of fully digital beamforming.

\bibliographystyle{IEEEtran}
\bibliography{bibliography}

\begin{thebibliography}{1}
\providecommand{\url}[1]{#1}
\csname url@samestyle\endcsname
\providecommand{\newblock}{\relax}
\providecommand{\bibinfo}[2]{#2}
\providecommand{\BIBentrySTDinterwordspacing}{\spaceskip=0pt\relax}
\providecommand{\BIBentryALTinterwordstretchfactor}{4}
\providecommand{\BIBentryALTinterwordspacing}{\spaceskip=\fontdimen2\font plus
\BIBentryALTinterwordstretchfactor\fontdimen3\font minus
  \fontdimen4\font\relax}
\providecommand{\BIBforeignlanguage}[2]{{%
\expandafter\ifx\csname l@#1\endcsname\relax
\typeout{** WARNING: IEEEtran.bst: No hyphenation pattern has been}%
\typeout{** loaded for the language `#1'. Using the pattern for}%
\typeout{** the default language instead.}%
\else
\language=\csname l@#1\endcsname
\fi
#2}}
\providecommand{\BIBdecl}{\relax}
\BIBdecl

\bibitem{Cisco}
CISCO, ``{Cisco Visual Networking Index: Forecast and Methodology,
  2016-2021},'' \emph{Cisco Public, 2017}.

\bibitem{Puglielli2016}
A.~Puglielli, A.~Townley, G.~LaCaille, V.~Milovanovic, P.~Lu, K.~Trotskovsky,
  A.~Whitcombe, N.~Narevsky, G.~Wright, T.~Courtade, E.~Alon, B.~Nikolic, and
  A.~M. Niknejad, ``{Design of Energy- and Cost-Efficient Massive MIMO
  Arrays},'' \emph{Proceedings of the IEEE}, vol. 104, no.~3, pp. 586--606,
  March 2016.

\bibitem{Sun2017}
Y.~Sun and C.~Qi, ``{Weighted Sum-Rate Maximization for Analog Beamforming and
  Combining in Millimeter Wave Massive MIMO Communications},'' \emph{IEEE
  Communications Letters}, vol.~21, no.~8, pp. 1883--1886, Aug 2017.

\bibitem{sohrabi2016hybrid}
F.~Sohrabi and W.~Yu, ``{Hybrid Digital and Analog Beamforming Design for
  Large-Scale Antenna Arrays},'' \emph{IEEE Journal of Selected Topics in
  Signal Processing}, vol.~10, no.~3, pp. 501--513, 2016.

\bibitem{utschick2018hybrid}
W.~Utschick, C.~St{\"o}ckle, M.~Joham, and J.~Luo, ``{Hybrid lisa precoding for
  multiuser millimeter-wave communications},'' \emph{IEEE Transactions on
  Wireless Communications}, vol.~17, no.~2, pp. 752--765, 2018.

\bibitem{bolcskei2000performance}
H.~Bolcskei and A.~J. Paulraj, ``{Performance of space-time codes in the
  presence of spatial fading correlation},'' in \emph{Asilomar Conference on
  Signals, Systems, and Computers}, vol.~1.\hskip 1em plus 0.5em minus
  0.4em\relax IEEE; 1998, 2000, pp. 687--693.

\bibitem{telatar1999capacity}
E.~Telatar, ``{Capacity of multi-antenna Gaussian channels},'' \emph{European
  transactions on telecommunications}, vol.~10, no.~6, pp. 585--595, 1999.

\bibitem{xie2016overview}
H.~Xie, F.~Gao, and S.~Jin, ``An overview of low-rank channel estimation for
  massive mimo systems,'' \emph{IEEE Access}, vol.~4, pp. 7313--7321, 2016.

\end{thebibliography}

\end{document}